# Digital Logarithmic Airborne Gamma Ray Spectrometer

GuoQiang ZENG*, QingXian ZHANG, Chen Li, ChengJun Tan,
LiangQuan GE, Yi GU, Feng CHENG

(Chengdu University of Technology, Chengdu, Sichuan, 610051, China)

**Abstract:** A new digital logarithmic airborne gamma ray spectrometer is designed in this study. The spectrometer adopts a high-speed and high-accuracy logarithmic amplifier (LOG114) to amplify the pulse signal logarithmically and to improve the utilization of the ADC dynamic range, because the low-energy pulse signal has a larger gain than the high-energy pulse signal. The spectrometer can clearly distinguish the photopeaks at 239, 352, 583, and 609keV in the low-energy spectral sections after the energy calibration. The photopeak energy resolution of $^{137}$Cs improves to 6.75% from the original 7.8%. Furthermore, the energy resolution of three photopeaks, namely, K, U, and Th, is maintained, and the overall stability of the energy spectrum is increased through potassium peak spectrum stabilization. Thus, effectively measuring energy from 20keV to 10MeV is possible.

## 1. Introduction

Airborne gamma ray spectrometry is a radiometric method, with a gamma ray spectrometer system installed on an aircraft to measure the gamma ray spectra of surface rocks and ore, from which the total radioactivity (or the intensity of radioactivity) and the contents of potassium, uranium and thorium are determined [1, 2]. The airborne gamma ray spectrometry method is also highly appropriate for large scale environmental surveys of areas of potentially contaminated ground, as well as for studies of the environment of nuclear sites for reference purposes [3].

Gamma-ray spectrometer (GRS) has been proven a kind of powerful instrument for remote measuring the abundance of chemical elements, like C, O, Mg, Al, Si, K, Ca, Fe, Th, and U on the planetary surface. The GRS onboard Chang'E-1 satellite (a lunar polar orbiter at the altitude 200km and in operation for one year) is conceived to map the lunar surface in the Chinese first lunar mission in 2007. The detector of GRS can detect the gamma rays of the moon at energies (300keV～10MeV).In this wide range spectrum measurement, the energy interval of x-axis is 22.14keV. Consider another reason of energy resolution at 662keV is 8.3%, the peak at 583keV and 727keV from Th lines overlapped with the peak at 609keV from element U [4].

In the above situation, the gamma ray energy in gamma ray spectrometry ranges up to 10MeV. The analog to digital converter (ADC) resolution and input signal range in the system are determined. Then, the pulse signals of the low-energy spectral section are significantly condensed under a linear amplifier to ensure that the pulse signals of the maximum energy 10MeV can be measured and analyzed. The low-energy spectral section among the collected spectrum is also condensed, and the photopeak energy resolution of the section significantly become worse, so is hard to satisfy the airborne gamma ray spectrometry survey [5, 6].

Segmented measuring is possibly a solution applied to guarantee the photopeak energy resolution of both high-energy and low-energy spectra simultaneously. In other words, it should use two spectrometers that have different amplifier gains to measure the high and low energy spectra individually. This method requires separate energy calibration and merges the two spectra together, which increases the hardware expense and data processing difficulty. Effective spectrum stabilization in the airborne gamma ray spectrometer measurement applies the

* Corresponding author: ZENG Guo-Qiang. E-mail: zgq@cdut.edu.cn

Supported by National Natural Science Foundation of China (Grant No.40904054) and National High Technology Research and Development Program 863 (Grant No. 2012AA061803).

natural potassium (K) photopeak as the characteristic photopeak. Segmented measuring method mentioned above could only stabilize the high energy spectra by natural potassium photopeak, but failed to the low energy spectra because K photopeak has been truncated in the low energy spectrometer. Large spectrum drift of the multi-crystal airborne gamma ray spectrometer measurement occurs, which significantly decreases the energy resolution of the photopeak of the multi-crystal merged spectrum [7].

The effective number of bits (ENOB) of ADC in the digital spectrometer decreases with the increase of the conversion rate and the input signal ranges of the digital spectrometer. The lower ENOB has lower resolution to pulse signals of the low energy gamma ray [8, 9]. For example, an ADC full scale input voltage range of 4V, which corresponds to the pulse signal amplitude of the 10MeV gamma ray, leads to pulse signal amplitude of only 8mV that corresponds to a 20keV gamma ray and is almost drowned out by the background noise. If a linear amplifier has to be used, the low-energy spectra are compressed and can not clearly show, it's signal dynamic range should be narrow. Using a logarithmic amplifier, the pulse signals from the low-energy radionuclides are greatly amplified, whereas the pulse signals from the high-energy radionuclides are less amplified. The photopeak interval of the high-energy rays among the produced logarithmic spectrum is consequently condensed and reduces the abundant blank spaces, whereas the photopeak interval of low-energy rays is widened. The result from the low-energy nuclides avoids wasting the blank spaces in the high-energy spectrum and significantly widens the low-energy spectrum, as well as recovers the energy resolution of the low-energy photopeak by maintaining the same energy resolution of the high-energy photopeak. This result takes full advantage of the channel address and the resolution of ADC.

Therefore, the logarithmic amplifying circuit is introduced in this paper to conduct the logarithmic calculation to the pulse signals. In doing so, this paper aims to condense the high-energy spectrum, widen the low-energy spectrum, and guarantee the energy resolution of both the high-energy and low-energy spectra.

## 2. System framework

Components of the digital logarithmic airborne gamma ray spectrometer are shown in Fig.1. The Ping-Pong buffer FIFO (First In First Out) memory coordinated with the quick interrupter built inside the ARM chip to achieve real-time parallel spectrum transmission and spectrum acquisition. The magnetic coupling serial port isolated the communication circuit through the ADM3251E and electrically isolated the digital spectrometer from the outside environment and serial port data communication. The maximum data transmission rate of the chip was 460kbps with an inbuilt DC–DC isolated power supply, which simplified the circuit design.

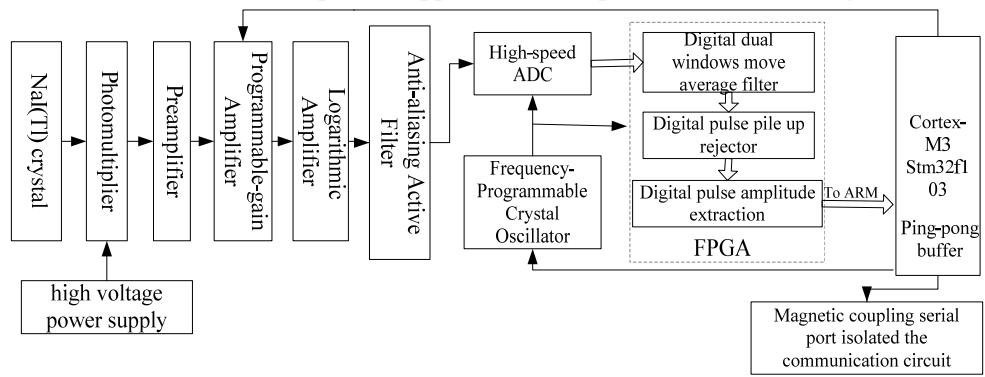

Fig.1. System framework of the digital logarithmic airborne gamma ray spectrometer

2.1 Design of digital pulse height analyzer

Since entering pulse signal is transformed through the logarithmic amplifier, instead of the usual fast-rising, slow decline of the double exponential signal, it cannot use digital trapezoidal shaper to analyze pulse height [10]. In this paper, the state machine based sampling method to achieve the maximum number pulse amplitude extraction. Is described as follows:

Construction of a width of 16 digital moving average window: average value of the first eight numbers defined as "*avg_forw*", average of the last eight numbers defined as "*avg_back*", "*ThUp*" defined as a variable indicates upper threshold of comparator, "*ThDn*" defined as a variable indicates lower threshold of comparator, "*base*" defines as a variable indicates baseline value, "*height*" defines as a variable indicates pulse amplitude.

**state1**: when expression (1) is true, initializes variable "*min*" and jumps to state2;

$$avg\_back < avg\_forw \qquad (1)$$

**state2**: read cycle variable of "*avg_forw*", and put the minimum value into "*min*"; when expression (2) is true, assigns variable "*min*" to variable "*base*", and initializes variable "*max*", jumps to state3;

$$avg\_forw > avg\_back + ThUp \qquad (2)$$

**state3**: read cycle variable of "*avg_forw*", and put the maximum value into "*max*"; when expression (3) is true, assigns variable "*height*" with result of "*max*" minus "*base*", there after output current variable "*height*" to ARM controller; jumps to state1;

$$avg\_forw <= avg\_back + ThDn \qquad (3)$$

From the foregoing, the method can not only signal filtering noise reduction, automatic tracking baseline, but also can distinguish two overlapped pulse signals. ADC sampling rate is higher, ability of pile-up correction will be more powerful [11]. In practical application of ADC sampling rate should be adjusted to the width of the moving average window in order to obtain the best filtering effect. By adjusting *ThUp* with *ThDn* value can filter out interfering signals, to guarantee that only the correct signal is processed by amplitude extraction.

This method does not do the exact dead time correction. Taking into account the ADC sampling rate of 40MHz, moving average window is 8 points, the signal rise times of 400ns, the system dead time is approximately 500ns. Hence, compensate dead time loss after each measurement according to the measuring time and total number of pulse signals [12].

2.2 Designed of preamplifier and programmable gain amplifier

Figure 2 shows the preamplifier and programmable gain amplifier designed in this system. Low bias current of high speed voltage amplifiers AD8065AR was selected to be preamplifier. AD8065AR configured for inverting amplifier mode, converted the current signal output from PMT to voltage signal. $D_1$, $D_2$ limited the voltage not exceeds ±0.6V of inverting input of $U_6$, protected the input of $U_6$ not to be damaged by surge voltage. Programmable gain amplifier was designed by using high speed multiplying DAC (AD5453YM[13]) that has 14bit resolution. This DAC with 10MHz multiplying bandwidth could meet the requirement of this system. Power supply of AD5453YRM is 3.3V, with STM32 control signal output level matching, and its Vref is connected with preamplifier output, Vref input impedance is approximately 10kΩ, can withstand ±12V maximum input signal. The input voltage signal is converted to a current signal and then converted current signal to voltage signal through the high-speed current op amp AD812AR. Figure 2, amplifier works on a logarithmic amplification mode when soldered $R_{27}$, removed $R_{28}$, signal of $U_{5B}$ through the logarithmic amplifier stage and input to ADC; when removed $R_{27}$, soldered $R_{28}$, signal of $U_{5B}$ direct output to the ADC, so work on linear amplification mode. After this preamplifier, the output pulse signal has peaking time [14] of 400ns, and falling time of 600ns.

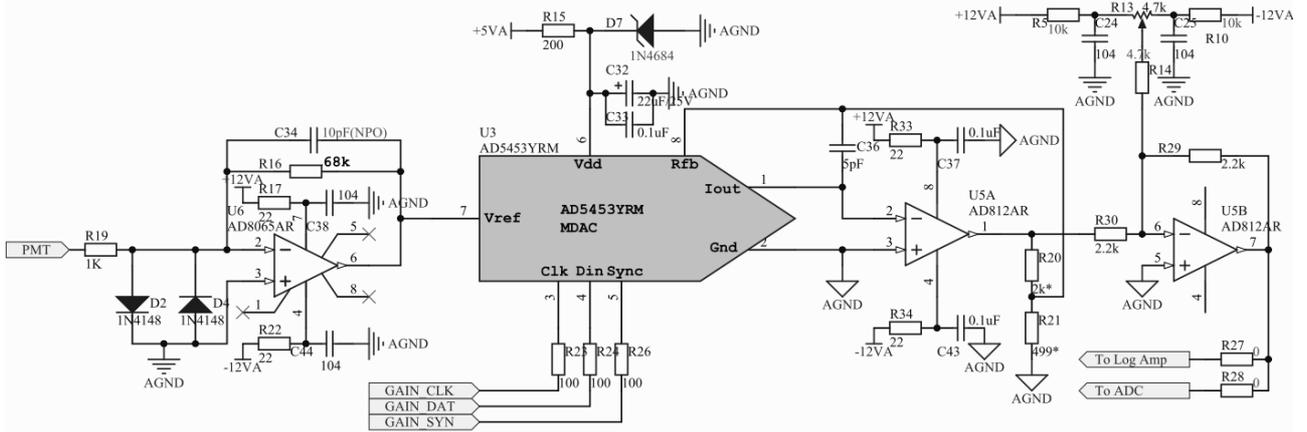

Fig.2. Schematics of preamplifier and programmable gain amplifier

Compared with the conventional digital airborne gamma ray spectrometer [15], the new digital logarithmic airborne gamma ray spectrometer adds a new logarithmic amplifier behind the preamplifier to logarithmically convert the pulse signals.

## 3. Design of logarithmic amplifier

The logarithmic amplifier designed in this paper is directly applied to amplify the pulse signal from preamplifier which hasn't been shaped, do not like that in [16] . Accuracy of the analyzer when pulse pile-up happened will deteriorate for the slow rising time of pulse signal. Hence, the important difference of this logarithmic amplifier is fast enough to preserve the short rising time of pulse signal from amplifier. The actual logarithmic amplifier is generally equipped with both linearity and logarithmic amplification, and is a linear amplifier with a larger gain for a weak input signal. However, the linear amplifier can become a logarithmic amplifier for strong input signals, and its gain will decrease with increasing input signal amplitude. The following are common types of logarithmic amplifiers: basic logarithmic amplifier, baseband logarithmic amplifier, and demodulating logarithmic amplifier.

The logarithmic amplifier used for precise pulse amplitude sampling has to have excellent DC precision and frequency response, which cannot be satisfied by common basic logarithmic amplifiers, because the pulse signals are double exponential signals that rise quickly but decay slowly. LOG114, a true logarithmic amplifier made by TI Company, has high DC precision, high speed, and high accuracy [17]. The bandwidth for small signals is larger than 10MHz, which meets the requirements of this research when $I_{ref}$ is at 10 µA. Fig.3 shows the internal functional framework of LOG114, which reveals that LOG114 logarithmically transforms the current inputs $I_1$ and $I_2$ into $V_{BE}$ voltage of the PN junction of the internal $Q_1$ and $Q_2$. $Q_1$ and $Q_2$ can overcome the DC error caused by the temperature drift because of the similar manufacturing technique and excellent matching of the audion. The $V_{BE}$ voltage output by $Q_1$ and $Q_2$ is differently amplified by the $A_3$ installed inside the LOG114, which further overcomes the DC error caused by the temperature drift. In addition, $A_4$ and $A_5$ can amplify the signal output of $A_3$. A 2.5V output reference source exists inside LOG114, which can be connected to Pin3 through the resistance to produce the reference current $I_2$. Based on the LOG114 datasheet [18], the output voltage of the operational amplifier $A_3$ is

$$V_{LOGOUT} = 0.375 \times \log_{10}(I_1 / I_2) + V_{CMIN} \qquad (4)$$

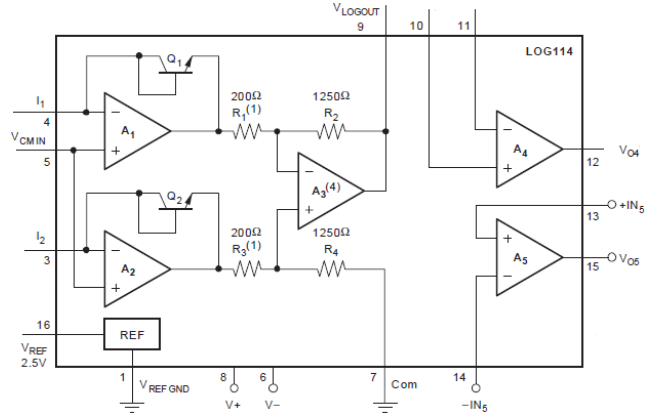

Fig.3. Internal functional framework of LOG114

The DC offset can be adjusted by altering $V_{CMIN}$. Given the bandwidth characteristics of the operational amplifiers $A_4$ and $A_5$ in LOG114 and the unsatisfying input offset voltage; this research applies the externally connected AD8065, which is a high-speed but low-noise operational amplifier, for signal amplification. The temperature coefficient of the internal reference source of LOG114 [16] is ±25ppm/°C. REF192ES is applied as the reference source because the temperature characteristic of the system worsens when the $I_2$ reference current produced by the internal reference source of LOG114 is used. The temperature coefficient of REF192ES [19] is 2 ppm/°C and its output voltage is 2.5V, which can increase the temperature stability of $I_2$. The practical design is shown in Fig.4.

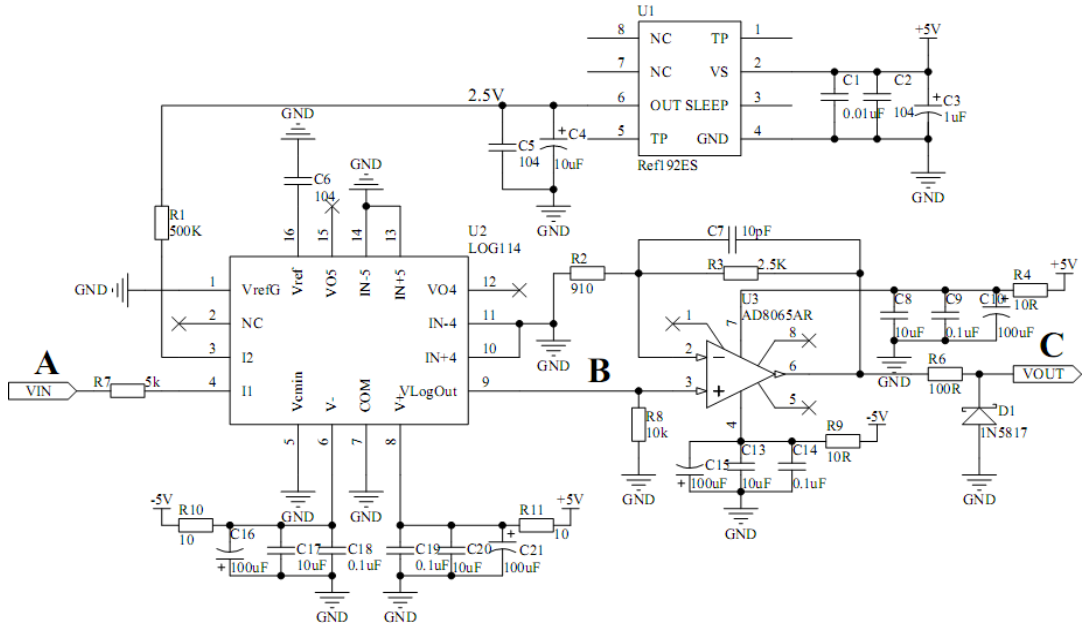

Fig.4. The circuit of the high-speed and high-accuracy logarithmic amplifier by LOG114

The pulse signals output by the preamplifier are voltage signals that have to be converted into current signals ($I_1$) when output to the logarithmic amplifier. A larger $I_1$ leads to a wider signal bandwidth based on the datasheet of LOG114. Therefore, $I_1$ is expected to be as large as possible. However, the fitting degree of the output logarithm decreases when $I_1>2mA$; thus, this study determines $I_{1MAX}=2mA$. The amplitude of pulse signal output by the preamplifier corresponding to the 10MeV gamma ray is assumed to be 10V; then, $R_7=10V/2mA=5k\Omega$. The DC level of the $V_{in}$ signal is 0V and the corresponding $I_1$ value is 0mA when the pulse signal is absent. Equation (4) states that the DC level of the output signal of LOG114 has a negative power supply (-5V). The DC level would be logarithmically amplified by LOG114 once pulse signals are present, and the output pulse would be at the B point in the circuit of Fig.4 (Fig.5 (a)). The positive amplitude part of the pulse signal in Fig.5 (a) is apparently the actual amplitude of the input pulse signal after the logarithmic calculation. Therefore, applying direct coupling for the connection of the output signals of LOG114 and next-stage circuit AD8065 is necessary.

The $D_1$ amplitude limiter on the post-amplifier output cuts the negative part of the outputting signal by using $R_6$ (Fig.4). A comparison between the original pulse signal and the quasi-triangular signal is shown in Fig.5 (b). The quasi-triangular signal is the output of the pulse signal, which is logarithmically amplified and shaped, and whose amplitude is limited by the operational amplifier, as shown in point C in the circuit of Fig.4. The output baseline is fixed at 0V after the amplitude is limited by $R_6$, $D_1$.

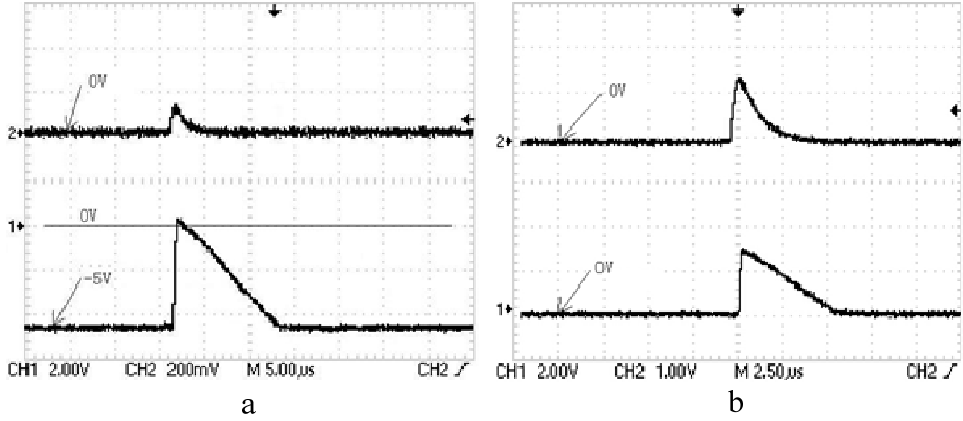

a                                                                  b

Fig.5. Signal of channel 2 in Fig (a) and Fig (b) comes from linear preamplifier's output; Signal of channel 1 in Fig (a) comes from point B of Fig.4; Signal of channel 1 in Fig (b) comes from point C of Fig.4

## 4. Relationship between reference current and spectrum

Equation (4) of the logarithmic amplifier output states that the output of the pulse signal through the logarithmic amplifier is related to the reference current $I_2$. The $V_{CMIN}$ used for adjusting the DC offset will be connected to the ground to maintain a simple subsequent design when the new DC offset is introduced. The new DC offset only has $\log_{10}(I_1/I_2)>0$ to ensure the output of the logarithmic amplifier $V_{LOGOUT}>0$, that is, $I_1>I_2$. Therefore, $I_2$ is used to set the minimum energy of the pulse signal. A larger $I_2$ or $I_1$ leads to a wider bandwidth of the DC signal of LOG114, a better high-frequency response, and a smaller pulse distortion based on the datasheet of LOG114. However, a larger $I_2$ would surely increase the lower energy threshold of the pulse signal, which necessitates setting an appropriate $I_2$. Only the value of $I_2$ is changed to obtain the optimal $I_2$. The spectra under the different $I_2$ are measured for comparison. Fig.6 shows that when $I_2$ is valued too high, the low-energy part of the spectrum is cut. The low-energy part can also be completely measured when $I_2$ is valued at 5μA. Therefore, when $I_2$=5μA and $R_1$=2.5V/5μA=500kΩ, the small signal bandwidth of $A_1$ is about 10MHz.

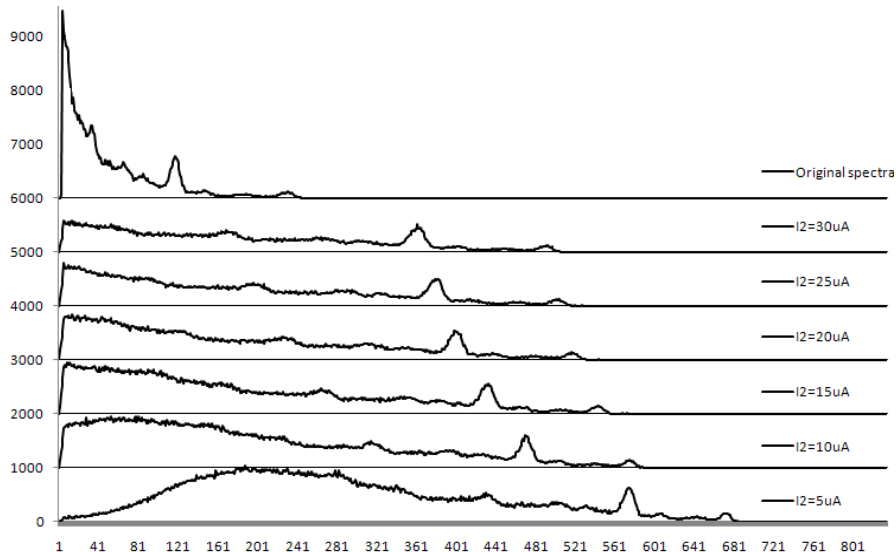

Fig.6. Natural gamma ray spectra measured with different reference currents

The output of logarithmic amplifier ($V_{LOGOUT}=\log_{10}(I_1/I_2)$) is inversely proportional to the $I_2$ current. Hence,

the smaller $I_2$ current, the wider of natural gamma ray spectrum.

## 5. Energy calibration of logarithmic spectrum

Energy calibration is essential before calculating the peak area and the content of the obtained logarithmic spectrum [20, 21]. The energy characteristic of the obtained spectrum is logarithmic as the pulse signals are processed with logarithmic transformation. Next, discussing the energy calibration of the spectrum that takes the logarithmic characteristic as its horizontal axis is necessary. The maximum voltage for the preamplifier to the output pulse signal is assumed as $V$ and the photon energy is $E$. Given that the preamplifier is a linear system, where $a$ and $b$ are constants, $V$ can be determined as

$$V = a \times E + b \quad (5)$$

The reference voltage of ADC is $V_{REF}$; the post amplification of logarithmic amplifier is $G$; and the corresponding channel address after the ADC conversion is $ch$ when the output pulse signals of the preamplifier enter the ADC after being logarithmically amplified by the logarithmic amplifier and if the resolution of the ADC is $2^n$. This result can be expressed as

$$\frac{G \times \log_{10}(V/R/I_2) \times 2^n}{V_{REF}} = ch \quad (6)$$

Where $R$ is the input resistance to convert the input voltage signals into current signals and $I_2$ is the reference current of LOG114, both of which are constants. Therefore, $I_2/R=K$. Equation (6) can be transformed as

$$\log_{10}(\frac{V}{K}) = \frac{V_{REF} \times ch}{G \times 2^n} \quad (7)$$

Then,

$$V = K \times e^{\frac{V_{REF} \times ch}{G \times 2^n}} \quad (8)$$

Equation (5) is substituted into Equation (7). Then,

$$a \times E + b = K \times e^{\frac{V_{REF} \times ch}{G \times 2^n}} \quad (9)$$

Finally,

$$e = \frac{K}{a} e^{\frac{V_{REF} \times ch}{G \times 2^n}} - \frac{b}{a} \quad (10)$$

Equation (10) reveals that an exponential functional relationship exists between the channel address and the ray energy. Therefore, the equation can be set as follows:

$$A_1 = K/a \quad (11)$$

$$A_2 = G \times 2^n / V_{REF} \quad (12)$$

$$A_3 = -b/a \quad (13)$$

Equations (11), (12), and (13) are substituted into Equation (10). Then,

$$e = A_1 \times e^{ch/A_2} + A_3 \quad (14)$$

$A_1$, $A_2$, and $A_3$ can be obtained through the exponential functional curve fitting on the energy of several groups of characteristic peaks and channel addresses. The substitution accomplishes the energy calibration of the logarithmic spectrum. Fig.8 shows the practical measured energy calibration curve, which has a reference current of 5μA. The calibration source of $^{241}$Am, $^{212}$Pb, $^{220}$Pb, $^{214}$Bi, $^{226}$Ra, $^{238}$Th, $^{137}$Cs, $^{40}$K and $^{208}$Tl are selected.

The energy calibration is calculated using Origin software (Fig.7), which obtains the calculation fitting degree of 0.999925 and the calculation result.

$$E = 14.561487 \times e^{0.006522ch} \quad (15)$$

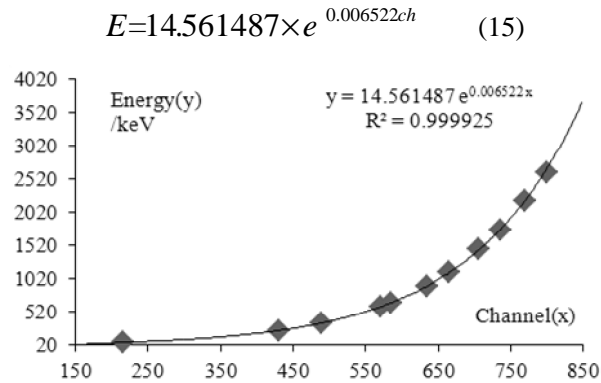

Fig.7. Energy Calibration Curve

The spectrum after the energy calibration based on the calculation result of Equation (15) is shown in Fig.8(b). Fig.8 (b) shows that the photopeaks at 239, 352, 583, and 609keV in the low-energy section can be obviously distinguished. Furthermore, the photopeak width of K, U, and Th are determined based on the energy interval selection approach of the airborne gamma ray spectrometry survey conducted by the International Atomic Energy Agency (IAEA)[21], as shown in Table 1.

Table.1 Recommended windows for natural radioelement mapping (IAEA, 1991)

| Window name | Isotope used | Gamma-ray energy /keV | Energy window /keV |
|---|---|---|---|
| Potassium | $^{40}$K | 1460 | 1370–1570 |
| Bismuth | $^{214}$Bi | 1760 | 1660–1860 |
| Thorium | $^{208}$Tl | 2615 | 2410–2810 |
| Total count | – | - | 410–2810 |
| Cosmic | – | - | > 3000 |

The energy interval division approach established by IAEA is also applicable on the logarithmic spectrometer, which achieves no overlapped energy intervals in the characteristic spectrum with all photopeaks independent from each other. Therefore, the application of the logarithmic calculation does not affect the high-energy gamma ray spectrum but widens the low-energy gamma ray spectrum, which is beneficial for the utilization and observation of the low-energy gamma ray spectrum and increases the utilization rate of the spectral data.

Figure 8 a, b show natural gamma ray spectra with energy range of 20keV～10MeV measured with digital airborne gamma ray spectrometer work in linear and logarithmic mode respectively. The measurement is carried out by using a 4inch * 4inch * 16inch size of NaI(Tl) crystals of $^{137}$Cs intrinsic energy resolution of 7.8% produced by Saint-Gobain Inc.

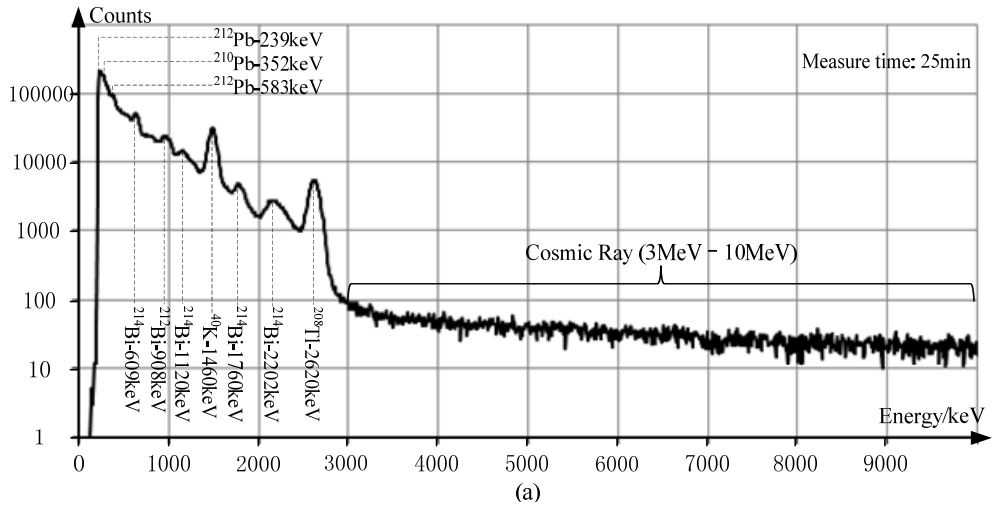

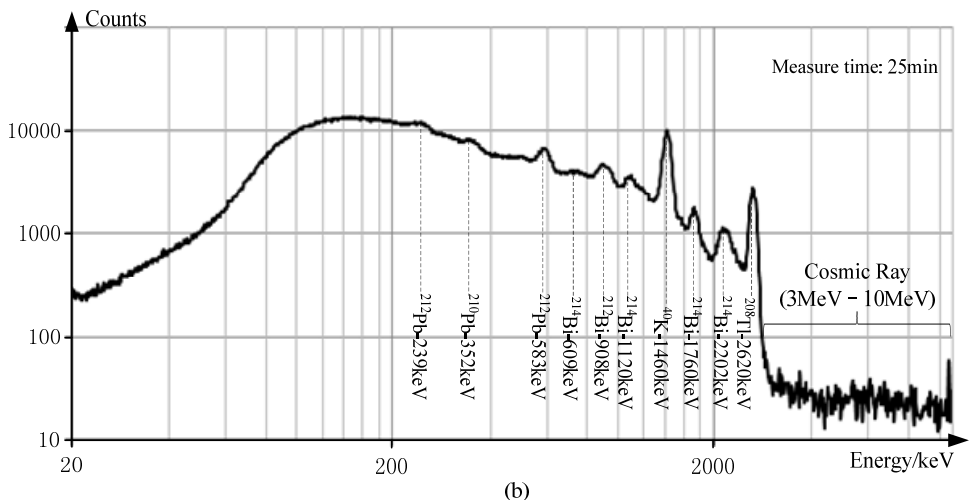

Fig.8 Contrast of natural gamma ray spectra with energy range of 20keV～10MeV measured in the same conditions; (a)Natural gamma ray spectrum measured in linear mode; (b) natural gamma ray spectrum measured in logarithmic mode.

## 6. Energy resolution comparison

A comparative measurement under the same conditions is conducted to distinguish the energy resolution differences between the digital logarithmic airborne gamma ray spectrometer and the conventional airborne gamma ray spectrometer. The measurements make use of the natural gamma ray spectrum and the $^{137}$Cs source spectrum for the same period using the same NaI(Tl) crystal detector and $^{137}$Cs source at the same location. Intrinsic energy resolution of the NaI(Tl) crystal for $^{137}$Cs is 6.6%.

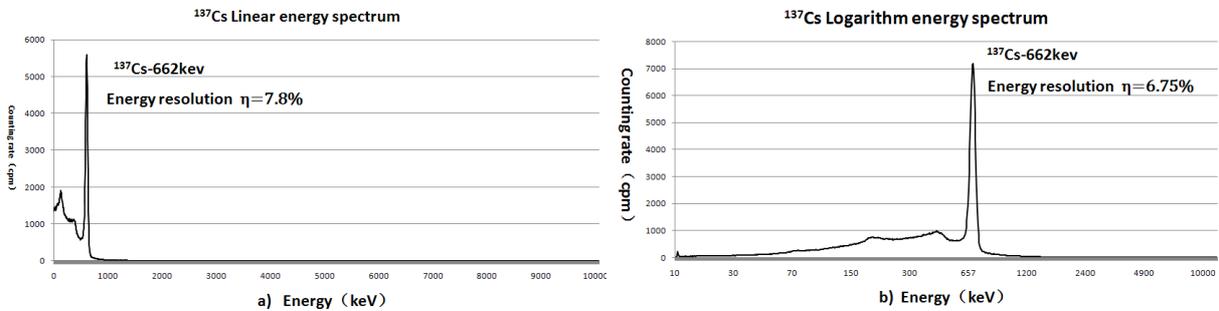

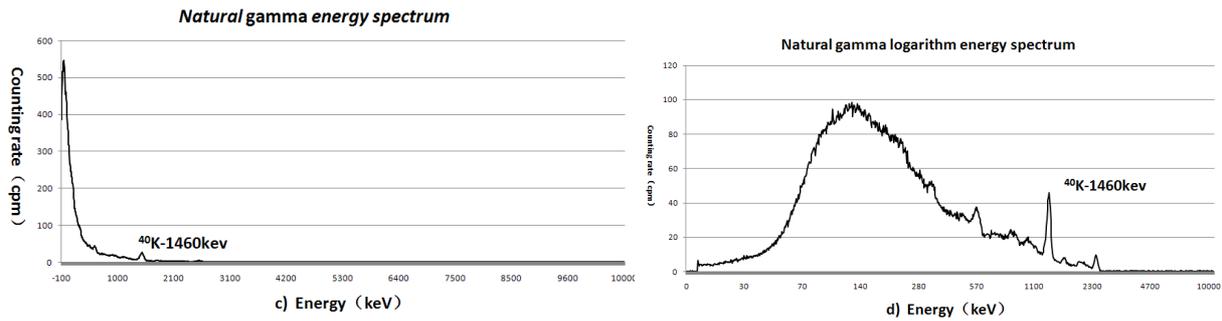

Fig.9. Comparison between the linear spectrum and the logarithmic spectrum

a) Linear spectrum of $^{137}$Cs      b) Logarithmic spectrum of $^{137}$Cs

c) Linear spectrum of natural gamma ray      d) Logarithmic spectrum of natural gamma ray

Two kinds of gamma ray spectrometer were designed to test. One of them has a logarithmic amplifier show in Fig.1, and the other one has similar components show in Fig.1 but remove logarithmic amplifier. The linear spectrum is shown in Fig.9(a). The logarithmic amplifier is used with the same NaI(Tl) crystal detector to measure the $^{137}$Cs source at the same location, which spectrum is shown in Fig.9(b). Similarly, the natural gamma spectrum is measured by the linear gamma ray spectrometer shown in Fig.9(c) and the logarithmic one is shown in Fig.9(d).

The energy resolutions ($DE(FWHM)/E$ ) of the linear spectra and the logarithmic spectra as shown in Table 2. Table 2 indicates that the energy resolution is related to the spectrum expansion. The energy resolution improves when the channel address corresponding to the photopeak of source is expanded by the logarithmic spectrum, but decreases when the channel address corresponding to the photopeak of source is condensed in the logarithmic spectrum.

Table.2. Comparison between high- and low-energy resolutions

| Element name | Isotope used | Gamma-ray energy /keV | Linear energy spectrum resolution | Linear channel range | Logarithm energy spectrum resolution | Logarithm channel range |
|---|---|---|---|---|---|---|
| Cesium | $^{137}$Cs | 662 | 7.8% | 71–80 | 6.75% | 508–528 |
| Potassium | $^{40}$K | 1460 | 6.12% | 157–169 | 5.56% | 616–634 |
| Thorium | $^{208}$Tl | 2615 | 3.66% | 264–281 | 4.2% | 702–712 |

Therefore, after being logarithmically amplified (LOG114), the digital logarithmic airborne gamma ray spectrometer can increase the energy resolution of its low-energy section by 1.05% and basically maintain the energy resolution of its high-energy section.

## 7. Conclusion

This paper describes a digital logarithmic airborne gamma ray spectrometer. This logarithmic spectrometer could maintain good resolution for both of low and high energy nuclides. Applying this logarithmic spectrometer as the energy of gamma rays being higher than 10MeV, two key factors should be noted: one is to choose the detector that has higher detecting efficiency for high energy gamma rays; the other is to complete efficiency calibration in the whole detecting energy range. Considering these two factors, the better option is to detect high energy gamma ray radiation source by using this logarithmic spectrometer. It should be known that we didn't consider the anti-radiation effect of this logarithmic spectrometer. Because airborne gamma ray spectrometry is environmental radiation detecting device, and its radiation dose is very low. Therefore, there is no need to consider the anti-radiation effects of electronic components. This logarithmic spectrometer can't be used in the environment where there has high radio activation because of the lack of overall anti-radiation evaluation. Obviously, the most effective way to improve the ability of resisting radiation is to carry out lead shielding to all of the electronic units. The practical measurement proves that this system can satisfy the wide energy range measurement of a multi-crystal airborne gamma ray spectrometry survey system.